# On linkage bias-correction for estimators using iterated bootstraps

Siu-Ming Tam[a], Min Wang[b], Alicia Rambaldi[c], Dehua Tao[d]

[a]No affliliation; [b]Faculty of Science and Technology, University of Canberra and School of Systems and Computing, University of New South Wales, Canberra, Australia; [c]School of Economics, University of Queensland, Queensland, Australia; [d]Research School of Finance, Actuarial Studies & Statistics, Australian National University, Canberra, Australia

**Contact** – *Siu-Ming Tam, stattam@gmail.com, Tam Data Advisory, Australia.*

**Abstract**

By amalgamating data from disparate sources, the resulting integrated dataset becomes a valuable resource for statistical analysis. In probabilistic record linkage, the effectiveness of such integration relies on the availability of linkage variables free from errors. Where this is lacking, the linked data set would suffer from linkage errors and the resultant analyses, linkage bias. This paper proposes a methodology leveraging the bootstrap technique to devise linkage bias-corrected estimators. Additionally, it introduces a test to assess whether increasing the number of bootstrap iterations meaningfully reduces linkage bias or merely inflates variance without further improving accuracy. An application of these methodologies is demonstrated through the analysis of a simulated dataset featuring hormone information, along with a dataset obtained from linking two data sets from the Australian Bureau of Statistics' labour mobility surveys.

1. **Introduction**

Confronted with diminishing budgets and escalating data collection costs exacerbated by factors such as declining survey response rates and a heightened demand for more extensive and frequent official statistics, national statistical offices (NSOs) are increasingly exploring data integration as a viable alternative to traditional direct data collection methods. The United Nations Economic Commission for Europe defines data integration as "the activity when at least two different sources of data are combined into a dataset. This dataset can be one that already exists in the statistical system or ones that are external sources (e.g., administrative dataset acquired from an owner of administrative registers or web-scraped information from a publicly available website)" (UNECE 2017).

Unless there are direct and accurate identifiers, e.g. names, addresses, company registration numbers etc. which enable the use of deterministic linking processes, probabilistic linking techniques have customarily been used for record linkage in official statistics. The methodology described in the seminal paper in Fellegi and Sunter (FS) (Fellegi and Sunter, 1969) has, since its publication, become the "gold" standard for probabilistic record linkage in NSOs.

Probability linking commences with the creation of an agreement matrix, $\Gamma = (\gamma_{11}, \gamma_{12}, ....., \gamma_{21}, \gamma_{22}, ...., \gamma_{ij}, ....\gamma_{n_A n_B})^T$ of order $(n_A \times n_B) \times L$ with row vectors, $\gamma_{ij}, i=1,...,n_A, j=1,...,n_B$, of order $1 \times L$, where $n_A, n_B$ and $L$ denote the size of data source A, data source B, and the number of linking variables used for record linkage respectively. In turn, $\boldsymbol{\gamma}_{ij} = (\gamma_{ij1},....,\gamma_{ijL})$ consists of $L$ 1's and 0's, where 1 denotes a match of a linking variable between unit $i \in A$ and unit $j \in B$, and 0 otherwise, for each of the $L$ linking variables. Using the EM algorithm (Samuels, 2012; Winkler, 2000), the quantities, $\hat{\boldsymbol{m}}^T = (\hat{m}_1,...,\hat{m}_L)$ and $\hat{\boldsymbol{u}}^T = (\hat{u}_1,...,\hat{u}_L)$, are EM estimates of $m_l = \Pr(\gamma_{ijl} = 1 | \gamma_{ij} \in M)$, and $u_l = \Pr(\gamma_{ijl} = 1 | \gamma_{ij} \in U)$ based on $\Gamma$, where $M$ and $U$ denote the unobserved set of true matched and true unmatched record pairs respectively and $i = 1,...,n_A, j = 1,...,n_B, l = 1,...,L$. Assuming further that the linking variables are statistically independent, the FS weight for the record pair $(i, j)$ is:

$$w_{ij} = \sum_{l=1}^{L} \{\gamma_{ijl} \log(\hat{m}_l) + (1-\gamma_{ijl})\log(1-\hat{m}_l) - \gamma_{ijl}\log(\hat{u}_l) - (1-\gamma_{ijl})\log(1-\hat{u}_l)\}.$$

The FS weights are used to determine the optimal linkage rule which, given a pre-determined false positive rate, $\mu$, and a false negative rate, $\lambda$, provides the cut-offs, $w_M$ and $w_U$ such that the record pair (i, j) belongs to $M$ if $w_{ij} \geq w_M$ or to $U$ if $w_{ij} \leq w_U$ and the number of undetermined record pairs to be resolved by a manual process is the minimum. It is easily seen that $w_M$ is inversely proportional to $\mu$. In other words, a higher tolerance for false positives will see a lower value of $w_M$. On the other hand, a higher tolerance for false negatives will see a higher value of $w_U$. For a more detailed account of probabilistic record linkage and the FS methodology, refer to, for example, Fellegi and Sunter (1969); Jaro (1989); Larsen and Rubin (2001), Sayers et al. (2016) and Winkler (1989, 1995). In this paper, we assume that there is no missing data with the linking variables. Extension of the case with missing linking variables can be addressed by extending $\gamma_{ijl}$ from a 1/0 variable to a -1/1/0, modifying the EM algorithm and the FS weights accordingly (Samuels, 2012).

The key assumptions used in the FS methodology for record linkage are (1) the linkage variables are statistically independent; and (2) there are no errors in the linkage variables, i.e. $\gamma_{ijl}$ are observed without error, which would otherwise adversely affect the computation of the FS weights, and the cut-offs $w_M$ and $w_U$. In practice, the number of matched and unmatched records pairs under the FS methodology is determined once $\mu$ and $\lambda$ are set, and the set of matched record pairs, $M$, and unmatched record pairs, $U$, will be obtained. However, if $\gamma_{ijl}$ is observed with error, the resultant sets, $M'$ and $U'$ will not be the same as $M$ and $U$ and statistical analysis based on $M'$ would therefore be subject to errors, hereinafter referred to as linkage errors. As an example, if the target variable, $y$, comes from source A, and the auxiliary variable, $x$, comes from source B, linkage error may result in $y_i$ incorrectly linked to $x_{\rho(i)}$ instead of $x_i$ and the resultant cross product sum $\sum_{i=1}^{n} y_i x_{\rho(i)}$, instead of $\sum_{i=1}^{n} y_i x_i$, has a linkage error bias when used to estimate the population cross product, $\sum_{1}^{N} x_i y_i$, where $N$ denotes the size of the finite population.

Instances in which $\gamma_{ijl}$'s are observed with error arise as a result of imperfect information in the linking variables which might include inconsistent or missing measurements, or modal errors from using different collection modes for the same variables between the same units in source A and source B. Methods for addressing linking errors based on parametric models have been developed in Lahiri and Larsen (2005); Larsen and Rubin (2001); Scheuren and Winkler (1993) and Tancredi and Lisco (2015) for regression analysis and in Chipperfield et al (2011) and Chipperfield and Chambers (2015) for contingency tables. Also, where the linking variables are correlated, statistically independence is violated and linking efficiency is affected. For examples of violation of the statistical independence assumption for business data, see Winkler (1985) and personal data, see Kelly (1986). Methods for addressing dependent linking variables are also discussed in Schurle (2015) and Yancy (2000).

This paper proposes methodologies for developing linkage bias-corrected estimators from integrated datasets generated using the FS algorithm. While our approaches do not specifically address the statistical dependence of linkage variables, this is not the main aim of this paper. We acknowledge that mitigating the violation of the statistical independence assumption of the linking variables may involve the removal or combination of correlated variables, or application of the methods of Schurle (2015) or Yancy (2000). The contributions of this paper to existing literature in correcting linkage bias encompass the following aspects: (1) the formulation of linkage-bias corrected estimators for parameters of unknown distribution functions (UDF); (2) a test to assess whether iterated bootstrapping has sufficiently reduced linkage bias to a statistically negligible level and (3) the establishment of bootstrap confidence intervals for bias-corrected estimators from the UDF. It should be noted that our methods, which are bootstrap based, are not applicable to certain estimators. For specific examples, refer to Athreya (1987) and Bickel and Freedman (1981).

The layout of this paper is as follows. Section 2 serves as a literature recap, focusing on the application of iterated bootstraps for estimating the bias of a parameter estimator, along with the construction of the bias-corrected estimator and their Percentile confidence intervals, as discussed I Efron (1987). Using the results of Section 2, Section 3 constructs linkage bias-corrected estimators and their Percentile confidence intervals from integrated data sets. In Section 4, we illustrate the methods by employing simulated hormone data in Efron and Tibshirani (1993). Additionally, we demonstrate the empirical effectiveness of these methods with an integrated dataset formed by merging two datasets from the Australian Bureau of Statistics. Section 5 provides our concluding remarks.

## 2. The iterated bootstrap

Suppose that we have a data set, $\chi$, of size $n$. Assume that the data points of $\chi$ are drawn independently from the UDF, $F$, and we are interested in estimating the parameters of $F$, denoted by $\theta(F)$, to emphasise its dependence on $F$. We use a statistic, $\hat{\theta}(\chi)$, compiled from the data set, $\chi$ to estimate $\theta(F)$. We are interested in removing the bias, if any, from $\hat{\theta}(\chi)$, to estimate $\theta(F)$, and computing the confidence interval of the bias-corrected estimator.

In this section, in order to provide some general results for iterated bootstraps, we assume that the bias of $\hat{\theta}(\chi)$ comes only from its functional form, e.g. an ML predictor using a training sample of $n$ $(y_i, x_i)$ record pairs to predict the out-of-sample population total of the target $y$ variable. In other words, we assume there is no linkage error. We then extend these results in the next section to the case where the bias comes not from the functional form but from linking errors.

**2.1 Bootstrap estimators and notation**

By definition, the functional form bias is given by $\boldsymbol{bias}_F = E_F\{\hat{\boldsymbol{\theta}}(\chi) - \boldsymbol{\theta}(F)\}$. Let $\chi^*$ be a bootstrap sample with the data points selected independently, with replacement, and with the same sample size as, $\chi$. We can compute a bootstrap estimate, $\hat{\boldsymbol{\theta}}^*(\chi^*)$, where $\hat{\boldsymbol{\theta}}^*(\chi^*)$ has exactly the same functional form as $\hat{\boldsymbol{\theta}}(\chi)$, but is computed over $\chi^*$ instead of $\chi$. Estimating $F$ by the empirical distribution, $\hat{F}_n$, by putting a probability mass $\frac{1}{n}$ on each of the data points in $\chi$, the "ideal" bootstrap estimate of $\boldsymbol{bias}_F$, denoted by $\hat{\boldsymbol{bias}}_{\hat{F}_n}$, is given by

$$\hat{\boldsymbol{bias}}_{\hat{F}_n} = E_{\hat{F}_n}\{\hat{\boldsymbol{\theta}}^*(\chi^*) - \hat{\boldsymbol{\theta}}(\chi)\} = E_{\hat{F}_n}\{\hat{\boldsymbol{\theta}}^*(\chi^*) - \boldsymbol{\theta}(\hat{F}_n)\}$$ (Efron and Tibshirani, 1993; Efron and Gong, 1983) where "ideal" is taken to mean that the estimated bias is computed over all possible samples that may be selected from $\hat{F}_n$. To save notation, we shall henceforth write $\hat{F}$ instead of $\hat{F}_n$. unless otherwise stated.

It can be shown that the number of possible samples with a sample size of $n$ is $C_{n-1}^{2n-1}$ and is bigger than 1 billion for $n \geq 17$. Limited by time and resources, only a relatively small number, $B$, of samples can be drawn, and $\hat{\boldsymbol{bias}}_{\hat{F}}$ is accordingly approximated by $\hat{bias}_B$, computed over the $B$, samples, $\chi_1^*,...,\chi_B^*$, sampled from $\hat{F}$, i.e. $\hat{bias}_B = \frac{1}{B}\sum_{b=1}^{B}\{\hat{\theta}_b^*(\chi_b^*) - \theta(\hat{F}_n)\}$ (Efron and Gong, 1983; Efron and

Tibshirani, 1986, 1993). Conditional on the observed data, the ideal bootstrap bias is

$\hat{bias}_{\hat{F}_n}$. Replacing it by $\hat{bias}_B$ yields a Monte Carlo error, $\hat{bias}_B - \hat{bias}_{\hat{F}_n}$ with order

$O_p((nB)^{-1/2})$ (Theorem 1(d), Chang and Hall, 2015). Denote $bias_F$ by $\tau(\chi)$ and let

$\tilde{\tau}^{(1)}(\chi)$ and $\tilde{\tau}^{(2)}(\chi)$ the bias estimated from the double bootstrap; also denote the

single and double bootstrap bias-corrected estimators by $\tilde{\theta}^{(1)}(\chi)$ and $\tilde{\theta}^{(2)}(\chi)$

respectively, then:

$$\tilde{\tau}^{(1)}(\chi) = \frac{1}{B}\sum_{b=1}^{B}\{\hat{\theta}_b^*(\chi_b^*) - \hat{\theta}(\chi)\} \qquad (1)$$

$$\tilde{\theta}^{(1)}(\chi) = \hat{\theta}(\chi) - \tilde{\tau}^{(1)}(\chi)$$
$$= \frac{1}{B}\sum_{b=1}^{B}\{2\hat{\theta}(\chi) - \hat{\theta}_b^*(\chi_b^*)\} \qquad (2)$$

$$\tilde{\tau}^{(2)}(\chi) = \frac{1}{BC}\sum_{b=1}^{B}\sum_{c=1}^{C}\{3\hat{\theta}_b^*(\chi_b^*) - \hat{\theta}_{bc}^{**}(\chi_{bc}^{**}) - 2\hat{\theta}(\chi)\} \qquad (3)$$

$$\tilde{\theta}^{(2)}(\chi) = \hat{\theta}(\chi) - \tilde{\tau}^{(2)}(\chi)$$
$$= \frac{1}{BC}\sum_{b=1}^{B}\sum_{c=1}^{C}\{3\hat{\theta}(\chi) - 3\hat{\theta}_b^*(\chi_b^*) + \hat{\theta}_{bc}^{**}(\chi_{bc}^{**})\} \qquad (4)$$

where $\hat{\theta}_{bc}^{**}(\chi_{bc}^{**})$ has the same functional form as $\hat{\theta}(\chi)$, but is computed over the $c^{th}$

sample selected independently and with replacement from the $b^{th}$ bootstrap sample

$\chi_b^*$, which we denote by $\chi_{bc}^{**}$ for $c = 1,...,C$. To save notation, and unless

otherwise indicated, we shall drop $\chi, \chi_b^*,$ and $\chi_{bc}^{**}$ from the estimators (1) to (4) in the sequel. More generally, if we denote the $k^{th}$ bootstrap estimate of $\theta$ by $\tilde{\theta}^{(k)}$ where $\tilde{\theta}^{(k-1)} = 0$ for $k=1$, then the following recursive relationship between the bootstrap estimate, bias and bias-corrected estimate hold (Hall and Martin, 1988):

$$\tilde{\theta}^{(k)} = \sum_{j=0}^{k} (-1)^j C_{j+1}^{k+1} \hat{\theta}^{*j} \qquad (5)$$

and

$$\tilde{\tau}^{(k)} = \hat{\theta} - \tilde{\theta}^{(k)}, \qquad (6)$$

where $\hat{\theta}^{*0} = \hat{\theta}, \hat{\theta}^{*1} = \hat{\theta}^*, \hat{\theta}^{*2} = \hat{\theta}^{**}$ etc..

## 2.2 How to determine $B$ and $k$ for higher-order bias reduction?

From the decomposition

$$\tilde{\theta}^{(k)} - \theta = \left[ E_{\hat{F}}(\tilde{\theta}^{(k)}) - \theta \right] + \left[ \tilde{\theta}^{(k)} - E_{\hat{F}}(\tilde{\theta}^{(k)}) \right] \qquad (7)$$

where the first square bracket term of (7) is the "ideal" bootstrap estimate of the bias, and the second square bracket term is the Monte Carlo error. Assuming that the random variables in the sample are mutually independent or independent within groups of roughly the same size, Theorem 1(a) of Chang and Hall (2015) shows that, under certain regularity conditions, and $B = B(n) \to \infty$ as $n \to \infty$, then $E_{\hat{F}}(\hat{\theta}^{(2)}) - \theta = O(n^{-3})$ and $E_{\hat{F}}(\hat{\theta}^{(1)}) - \theta = O(n^{-2})$, i.e. the double bootstrap provides a higher degree of accuracy in terms of bias correction. As shown later, this is at the cost of higher variance. Hall and Martin (1988) extended the results of Theorem 1(a) to $E_{\hat{F}}(\hat{\theta}^{(k)}) - \theta = O(n^{-(k+1)}), k \geq 3$. Theorem 1(d) of Chang and Hall (2015) also showed that the Monte Carlo error in (7) is $O_p((nB)^{-1/2})$, for $k = 1, 2$. Because $\left[ E_{\hat{F}}(\tilde{\theta}^{(k)}) - \theta \right]$ is $O(n^{-(k+1)})$ and $\left[ \tilde{\theta}^{(k)} - E_{\hat{F}}(\tilde{\theta}^{(k)}) \right]$ is $O_p((nB)^{-1/2})$, we need $B >> n^{2k+1}$ so that the higher-order bias reduction at iteration $k$ remains detectable in the presence of Monte Carlo error.

Let $\tilde{d}_k = \tilde{\tau}^{(k-1)} - \tilde{\tau}^{(k)} = \tilde{\theta}^{(k)} - \tilde{\theta}^{(k-1)} = \Delta(\tilde{\theta}^{(k)})$ (with $\tilde{\tau}^{(0)}$ defined as the null vector), then from (7),

$$\Delta(\tilde{\theta}^{(k)}) = \left[ E_{\hat{F}}(\tilde{\theta}^{(k)}) - E_{\hat{F}}(\tilde{\theta}^{(k-1)}) \right] + \left[ \{\tilde{\theta}^{(k)} - E_{\hat{F}}(\tilde{\theta}^{(k)})\} - \{\tilde{\theta}^{(k-1)} - E_{\hat{F}}(\tilde{\theta}^{(k-1)})\} \right] \quad (8)$$

where the first term of (8) is the "ideal" k-level bias reduction term, and each of the two curly bracket term is the Monte Carlo error. Let

$\Delta(\boldsymbol{a}_k) = (\Delta a_{k,i}) = \tilde{\theta}^{(k)} - E_{\hat{F}}(\tilde{\theta}^{(k)})$ where $i = 1,..,\dim(\boldsymbol{a}_k) = p$. From Theorem 1(d) of Chang and Hall (2015), for $k = 1, 2$, $a_{k,i}$ has an asymptotic normal distribution with mean zero, and $Var(a_{k,i}) = O(1/(nB)) = \dfrac{C}{nB}$ which notably is independent of $k$, where $C$ is a constant. For $i = 1,..., p, a_{k,i}$ is the "ideal" bias at the $k^{th}$ iterated bootstrap, and $b_{k,i}$ is the Monte Carlo error due to the use of a finite number of bootstrap samples. Let $\Delta(\boldsymbol{b}_k) = (\Delta b_{k,i}) = \left[ E_{\hat{F}}(\tilde{\theta}^{(k)}) - E_{\hat{F}}(\tilde{\theta}^{(k-1)}) \right]$. It follows from Theorem 1(a) of Chang and Hall (2015) that $b_{k,i} = O(n^{-(k+1)}) - O(n^{-k}) = O(n^{-k})$.

Given this, to test if the $k^{th}$ iterated bootstrap has any detectable effect on bias reduction, i.e. whether $b_{k,1} - b_{k-1,i}$ is statistically significant from zero, the bootstrap sample size $B_i$ must satisfy $\sqrt{\dfrac{2C(1-\rho_i)}{nB}} < \dfrac{1}{n^k}$ or

$$B_i > 2C(1-\rho_i)n^{2k-1}, \text{ and}$$
$$B = \max\{B_i\}_1^p, \qquad (9)$$

where $\rho_i$ is the correlation coefficient between $a_{k,i}$ and $a_{k-1,i}$, where common random numbers are used. Clearly condition (9) is not practical even for small $n$ or $k$. Instead, we propose the following method for testing.

Let $\tilde{d}_k = b_k + a_k = (\Delta(\tilde{\theta}_i^{(k)}))$. Then, for $i=1,..,p$,

$$\begin{aligned}
Var(\Delta(\tilde{\theta}_i^{(k)})) &= Var(E_{\tilde{F}}(\Delta(\tilde{\theta}_i^{(k)}) \mid \chi)) + E_{\hat{F}}(Var(\Delta(\tilde{\theta}_i^{(k)}) \mid \chi)) \\
&= Var(\Delta(b_{k,i})) + E_{\hat{F}}[\dfrac{2C(1-\rho_i)}{nB_i}] \\
&= Var(\Delta(b_{k,i})) + \dfrac{2C(1-\rho_i)}{nB_i} \\
&= V_{1,i}^{(k)} + \dfrac{V_{2,i}^{(k)}}{B_i} \\
&= V_{1,i}^{(k)} (1+\eta_i^2), \qquad (10)
\end{aligned}$$

where $\eta_i^{(k)2} = \dfrac{V_{2,i}^{(k)}}{B_i V_{1,i}^{(k)}}$ and $B_i$ represents the sample size chosen for the $i^{th}$ parameter of $\theta$. Treating $\chi$ as a realization from a superpopulation model, $V_{1,i}$ can be seen as the sampling variance of the bias, and $V_{2,i}$ is the Monte Carlo variance.

Our goal is to find the smallest $k$ such that the statistic, $\frac{\Delta(b_{k,i})}{\sqrt{V_{1,i}}}$, which has a mean of 0 and variance of 1, is not significantly different from zero – then additional bias removed by using the $k^{th}$ iterated bootstrap is not detectable. Because $\Delta(b_{k,i})$ is unobserved, and from

$$\frac{\Delta(\tilde{\theta}_i^{(k)})}{\sqrt{V_{1,i}^{(k)}}} = \frac{\Delta(b_{k,i})}{\sqrt{V_{1,i}^{(k)}}} + \frac{\Delta(a_{k,i})}{\sqrt{V_{1,i}^{(k)}}}, \tag{11}$$

we have to turn to testing the studentized test statistic (Hall, 1986), $\frac{\Delta(\tilde{\theta}_i^{(k)})}{\sqrt{V_{1,i}^{(k)}}}$, which has a mean zero, but a variance of $1+\eta_i^{(k)2}$. So that the test of significance of $\frac{\Delta(b_{k,i})}{\sqrt{V_{1,i}^{(k)}}}$ by using the observed $T_i^{(k)} = \frac{\Delta(\tilde{\theta}_i^{(k)})}{\sqrt{V_{1,i}^{(k)}}}$ as a proxy is not distorted, and noting the confidence interval of $T_i^{(k)} = \frac{\Delta(\tilde{\theta}_i^{(k)})}{\sqrt{V_{1,i}^{(k)}}}$ will be correct to $O(\eta_i^{(k)2})$, we can set $\eta_i^{(k)}$ to be a small $\eta_0$, and choose

$$B_i^{(k)} \geq \frac{V_{2,i}^{(k)}}{\eta_0^2 V_{1,i}^{(k)}}, B^{(k)} = \max_i \{B_i^{(k)}\}. \tag{12}$$

There is a trade-off between accuracy of the test statistic by setting $\eta_0$ to be close to zero, i.e. not stopping at $k$ too early (see next paragraph) and computational burden due to a bigger $B^{(k)}$.

We test to find the smallest $k$ such that $H_0 : \Delta(b_{k,i}) = 0$ cannot be rejected for all $i$ at 95% confidence level, i.e there is no detectable incremental bias reduction using $\tilde{\theta}_i^{(k)}$ over $\tilde{\theta}_i^{(k-1)}$ for all $i$ for this $k$. Under $H_0$, $E[\Delta(\tilde{\theta}_i^{(k)})] = 0$ and $Var(T_i^{(k)}) = 1 + \eta_i^{(k)2}$. The 95% percentile confidence interval of $T_i^{(k)}$ is $(q_{l,i}^{(k)}, q_{u,i}^{(k)})$ where $q_{l,i}^{(k)}, q_{u,i}^{(k)}$ are the 2.5% and 97.5% percentile of $T_i^{(h;k)}, h = 1,...,H$ constructed from $H$ outer samples. The hypothesis $H_0$ will be rejected if the 95% percentile interval
$$\text{CI}_i^{(k)} = (\Delta(\tilde{\theta}_i^{(k)}) + q_{l,i}^{(k)}\sqrt{V_{1,i}^{(k)}}, \Delta(\tilde{\theta}_i^{(k)}) + q_{u,i}^{(k)}\sqrt{V_{1,i}^{(k)}})$$ does not contain 0. So our task is the find the first $k$ such that $\text{CI}_i^{(k)}$ contains 0, indicating that the $(k-1)^{th}$ iterated bootstrap corrected estimator is as best as one can get statistically. Where this $k$ is found, we can stop the bias correction at $\tilde{\theta}_i^{(k-1)}$. For ease of reference, we shall henceforth name the test as CI test.

To estimate $Var(\Delta(\tilde{\theta}_i^{(k)}))$, $V_{1,i}^{(k)}$ and $V_{2,i}^{(k)}$ needed for computing the percentile confidence interval of $T_{k,i}$, we proceed as follows. For the outer bootstrap samples, $h = 1,...,H$, and inner bootstrap samples, $b = 1,...,B$, let $\Delta_i^{(h,b;k)} = \tilde{\theta}_i^{(h,b;k)} - \tilde{\theta}_i^{(h,b;k-1)}$, where $\tilde{\theta}_i^{(h,b;k)}, \tilde{\theta}_i^{(h,b;k-1)}$ represent the $k^{th}$ and $(k-1)^{th}$ iterated bootstrap corrected $i^{th}$ estimator from the nested $hb^{th}$ sample. We then have:

(i) $\quad V\hat{a}r(\Delta(\tilde{\theta}_i^{(k)})) = \dfrac{1}{H-1} \sum_{h=1}^{H} (\overline{\Delta}_i^{(h;k)} - \overline{\overline{\Delta}}_i^{(k)})^2$, where

$$\overline{\Delta}_i^{(h;k)} = \dfrac{1}{B} \sum_{b=1}^{B} \Delta_i^{(h,b;k)}, \overline{\overline{\Delta}}_i^{(k)} = \dfrac{1}{H} \sum_{h=1}^{H} \overline{\Delta}_i^{(h;k)}; \quad (13)$$

(ii) $\quad \hat{V}_{2,i}^{(k)} = \dfrac{1}{H} \sum_{h=1}^{H} \hat{V}_{2,i}^{(h;k)}$, where $\hat{V}_{2,i}^{(h;k)} = \dfrac{1}{B-1} \sum_{b=1}^{B} (\Delta_i^{(h,b;k)} - \overline{\Delta}_i^{(h;k)})^2$;

(14)

and

(iii) $\quad \hat{V}_{1,i}^{(k)} = V\hat{a}r(\Delta(\tilde{\theta}_i^{(k)})) - \dfrac{\hat{V}_{2,i}^{(k)}}{B_i^{(k)}}.$

(15)

To ensure that $\hat{V}_{1,i}^{(k)} > 0$, we initially choose $B_i^{(k)} = B_{i0}^{(k)}$ such that

$B_{i0}^{(k)} > \dfrac{\hat{V}_{2,i}^{(k)}}{V\hat{a}r(\Delta(\tilde{\theta}_i^{(k)}))}$. When then recompute $B_i^{(k)}$ using (12) and then make the final choice of $B_{i,k,\max} = \max\{B_i^{(k)}, B_{i0}^{(k)}\}$ to ensure that conditions (12) and (15) are satisfied. Computing $H$ bootstrap estimates of the variances in (13) to (15) and $T_{k,i}$ requires computing $\Delta(\tilde{\theta}_i^{(k)})$ $HxB_{i,k,\max}$ times, which can be computationally very burdensome, particular when $H$ and $B_{i,k,\max}$ are large. For (1), (2) and (3), we set $H$ and $B$ to be of moderate size, say 100 each to get reasonable estimates. However, for $T_{k,i}$, the computation for $HxB_{k,\max}$ bootstrap samples could be considerable. Inspired by Otsu and Rai (2017)(OR), we now outline a less labour and computationally intensive effort to draw the $H$ bootstrap samples. Note that all the estimators in this paper are of the form $\hat{\bar{\lambda}}_M = \frac{1}{M}\sum_{i=1}^{M}\lambda_i$ with suitable choices of $M$ and $\lambda_i$, where $\hat{\bar{\lambda}}_M$ is used to estimate some population parameter, $\bar{\lambda}$ ; for example,

$$\tilde{\tau}_i^{(1)}(\chi) = \frac{1}{B_{i,1,\max}}\sum_{b=1}^{B_{i,1,\max}}\left\{\hat{\theta}_{i,b}^* - \hat{\theta}_i\right\};$$

$$\tilde{\theta}_i^{(2)}(\chi) = \frac{1}{B_{i,2,\max}}\sum_{b=1}^{B_{i,2,\max}}\left[\frac{1}{C}\sum_{c=1}^{C}\left\{3\hat{\theta}_i - 3\hat{\theta}_{i,b}^* + \hat{\theta}_{i,bc}^{**}\right\}\right];$$

and

$$\begin{aligned}\tilde{\theta}_i^{(k-1)} - \tilde{\theta}_i^{(k)} &= \tilde{\tau}_i^{(k)} - \tilde{\tau}_i^{(k-1)} \\ &= \frac{1}{B_{i,2,\max}}\sum_{b=1}^{B_{i,2,\max}}\left[\frac{1}{C}\sum_{c=1}^{C}\left\{2\hat{\theta}_{i,b}^* - \hat{\theta}_{i,bc}^{**} - \hat{\theta}_i\right\}\right], k = 2 \text{ or} \\ &= \frac{1}{B_{i,3,\max}}\sum_{b=1}^{B_{i,3,\max}}\frac{1}{C}\left[\sum_{c=1}^{C}\frac{1}{D}\sum_{d=1}^{D}(3\hat{\theta}_{i,b}^* - 3\hat{\theta}_{i,bc}^{**} + \hat{\theta}_{i,bcd}^{***} - \hat{\theta}_i)\right], k = 3 \text{ etc..}\end{aligned} \quad (16)$$

Under their method, each of the $H$ bootstrap estimator can be created by taking a random sample of size $M$ independently and with replacement from $\{\{\lambda_1,...,\lambda_M\}\}$. In the numerical examples below, we set $H = 2,000$. In the sequel, we refer this method of bootstrap as OR. We can use a similar idea to compute the bootstrap test statistic, $T_i^{(k)} = \frac{\Delta(\tilde{\theta}_i^{(k)})}{\sqrt{V_{1,i}^{(k)}}}$, where $\Delta(\tilde{\theta}_i^{(k)})$ and $V_{1,i}^{(k)}$ are computed using (16) and, as a "plug in", (14) respectively.

In addition, it is worth noting that while the bias of $\tilde{\theta}^{(k)}$ may be smaller than $\tilde{\theta}^{(k-1)}$, they have larger variances. This highlights the role of bootstrapping in bias reduction within the broader bias-variance trade-off context. Ultimately, the choice between prioritizing bias reduction or variance control depends on the decision-maker's specific needs and priorities.

Finally, we agree with the comment of a referee that it is generally computational intensive to use bootstraps for the bias correction methodology outlined in this paper. Thankfully, the availability of Fastlink (Enamorado et al., 2019) and SpLink (MoJ, 2021), and open source Spark-enabled Python package, developed by the UK Ministry of Justice based on the ideas of Fastlnk, are efficient computational tools for linking large data sets. Together with the weighted bootstrap method of Otsu and Rai (2017), the computational burden is, though not completed eliminated, has been reduced.

## 3. Application to integrated data sets for linkage bias and functional form bias correction

We now consider that $\chi$ is an integrated data set of linked record pairs, being a subset of the Cartesian product $(n_A \times n_B)$ of all the possible record pairs between the units in data source A and data source B, and with the record pairs linked using the FS algorithm. Here $n_A$ and $n_B$ denote the sample size of A and B respectively. For integrated data sets, in addition to bias from the functional form of $\hat{\theta}$, bias arises from an improper constituted $\chi$ caused by linking error, i.e. $\gamma_{ijl}$ observed with error, or the incorrect assumption of statistical independence of the linking variables used in determining the weights in the FS algorithm. In this section, we limit our discussion to addressing linkage error bias only and assume that there is no dependence between the linking variables. This can be achieved by either eliminating or merging highly correlated variables or, alternatively, by using the methods outlined in Chipperfield et al. (2011) and Chipperfield et al. (2018). Classical bootstrap sampling does not eliminate linkage error because it selects data points that may already contain linkage bias. As a result, the resampled datasets inherit the same systematic errors present in the original data, preventing bias correction. We need a different method to select the bootstrap samples to correct for linkage error bias

Recalling the sample $\chi$ is constituted by linking the units in data source A with data source B based on the FS weights, which in turn is based on the agreement matrix, $\Gamma = (\gamma_{11}, \gamma_{12}, ...., \gamma_{21}, \gamma_{22}, ..., \gamma_{ij}, ....\gamma_{n_A n_B})^T$ where $\gamma_{ij} = (\gamma_{ij1}, ...., \gamma_{ijL})$ consists of $L$ 1's and 0's, with 1 denoting a match of a linking variable between unit $i \in A$ and $j \in B$, and 0 otherwise. We denote by $\Gamma_{X^M}$ the agreement matrix by stacking up the $\gamma_{ijl}^T$'s of linked record pairs, and $\Gamma_{X^U}$ the stacking up of $\gamma_{ijl}^T$'s remaining non-linked record pairs, where X denotes the set of possible record pairs between sources A and B, formed by the Cartesian product between the units of both sources. Without loss of generality, we sort the agreement matrix in a way such that $\Gamma = \begin{pmatrix} \Gamma_{X^M} \\ \Gamma_{X^U} \end{pmatrix}$. Because $\chi$ depends on $\Gamma$, formally we should write $\chi = \chi(\Gamma)$.

Following Chipperfield and Chambers (2015) and Chipperfield et al (2018), we consider $\gamma_{ij}$ as a random vector of variables, with $\gamma_{ijl} \sim Bern(\hat{m}_l)$ or $Bern(\hat{u}_l)$ if $\gamma_{ij} \in \Gamma_{X^M}$ or $\gamma_{ij} \in \Gamma_{X^U}$ respectively, for $i = 1,...,n_A, j = 1,..,n_B, l = 1,...,L$. Here, $\hat{m}_l$ and $\hat{u}_l$ denote the fixed probabilities estimated from the original linked dataset. In this formulation, $\gamma_{ijl}$ is treated as a realization of the joint Bernoulli distribution, given $\hat{m}_l$ and $\hat{u}_l$, which remain fixed throughout the analysis to isolate the variation caused by linkage errors. Similarly, $\Gamma_{X^M}$ is considered to be a realization of the joint Bernoulli distribution $f(\hat{m}_1,...,\hat{m}_L) = \Pi_{i=1}^{L} Bern(\hat{m}_i)$ and $\Gamma_{X^U}$ a realization of the joint

Bernoulli distribution $f(\hat{u}_1,...,\hat{u}_L) = \Pi_{i=1}^{L} Bern(\hat{u}_i)$. Accordingly, steps 1- 4 and steps 5-9 describe the procedures for creating single (parametric) bootstrap samples,

and double and triple bootstrap samples respectively:

(1) If $\gamma_{ij} \in \Gamma_{X^M}$, draw $\gamma_{ijl}^* \sim Bern(\hat{m}_l)$; if $\gamma_{ij} \in \Gamma_{X^U}$, draw $\gamma_{ijl}^* \sim Bern(\hat{u}_l)$, for

$i = 1,...,n_A, j = 1,..,n_B, l = 1...L$;

(2) Create $\Gamma^*$ by stacking up the $\gamma_{ijl}^*$;

(3) Compute the new FS weights based on $\Gamma^*$ to create a new integrated data set,

$\chi^* = \chi(\Gamma^*)$;

(4) Repeat steps (1) to (3) $B$ times to create $\chi_b^* = \chi(\Gamma_b^*)$ and compute $\hat{\theta}_b^*$,

$b = 1,..., B.$

Equally by sorting the agreement vectors appropriately, we can write

$\Gamma_b^* = \begin{pmatrix} \Gamma_{X_b^M}^* \\ \Gamma_{X_b^U}^* \end{pmatrix}$. Double (parametric) bootstraps samples are then created, for every

$\chi_b^* = \chi(\Gamma_b^*), b = 1,..., B,$, as follows:

(5) If $\gamma_{bij}^* \in \Gamma_{X_b^M}^*$, draw $\gamma_{bijl}^{**} \sim Bern(\hat{m}_l)$; if $\gamma_{bij}^* \in \Gamma_{X_b^U}^*$, draw $\gamma_{bijl}^{**} \sim Bern(\hat{u}_l)$, for

$i = 1,...,n_A, j = 1,..,n_B, l = 1...L$;

(6) Create $\Gamma^{**}$ by stacking up the $\gamma_{bijl}^{**}$;

(7) Calculate the new FS weights based on $\Gamma^{**}$ to create a new integrated data set,

$\chi^{**} = \chi(\Gamma^{**})$;

(8) Repeat steps (5) to (7) $C$ times to create $\chi_{bc}^{**} = \chi(\Gamma_{bc}^{**})$ and $\hat{\theta}_{bC}^{**}, c = 1,...,C$;

(9) Adapt steps (5) to (8) to $\chi_{bc}^{**}$ to create triple bootstrap samples $\chi_{bc}^{***} = \chi(\Gamma_{bcd}^{***})$ and compute $\hat{\theta}_{bcd}^{***}, d = 1,...,D$;

(10) Select $H$ samples using the OR method to compute the percentile confidence intervals for $\tilde{\theta}^{(k-1)}$ and $\tilde{\theta}^{(k)} - \tilde{\theta}^{(k-1)}$, $k = 2, 3$ etc..

(11) Apply the CI test for each parameter. Repeat steps (1) to (10) for each parameter that fails the test.

For ease of reference, shall refer the sample created using Steps 1 to 10 above as Agreement Matrix Bootstrap Integrated Sample (AMBIS), and the method to create the bootstrap pairs for matching as Agreement Matrix Bootstrap Pairs (AMBoP). Unlike Section 2, in which the classical (non-parametric) bootstrap samples are created by sampling the data points in $\chi$, in this section, AMBIS are created by sampling the row vectors of the agreement matrix $\Gamma$ using the two joint Bernoulli distributions, with the exception of estimating the variance terms in (12), (13) and (14), where the $H$ samples are generated by OR sampling. Note that in Step (5) above, we use $\hat{m}_l$ and $\hat{u}_l$ generated in Step (1) instead of re-estimation. This is because the bootstrap samples are generated by using $\hat{m}_l$ and $\hat{u}_l$ and recalculation will at best give the same values back. Once can also consider the AMBIS are created conditional on $\hat{m}_l$ and $\hat{u}_l$.

## 4 Simulation and Empirical Application

We illustrate our methods by providing a simulation (Section 4.1) and present an empirical example using a transition to work model with Australian data.

## 4.1 Simulated data illustration

In this section we use the hormone data of Efron and Tibshirani (1993, page 110). The data is reproduced in Appendix 1. We assume that the data set is constituted by linking without linking errors two data sources. This is used as the "true" dataset in this experiment. The response variable, $y_i$,, represents the amount of hormone in medical device $i$ and the explanatory variable, $x_i$, represents the number of hours the $i^{th}$ device was worn. The linear model $y_i = \beta_o + \beta_1 x_i + \varepsilon_i, i = 1,...,27$ is assumed by Efron and Tibshirani (1993) where the error terms $\varepsilon_i$ are distributed independently from an unknown distribution, $F$. The parameters for this data set are $\boldsymbol{\theta}(F) = (\beta_0, \beta_1)^T$. Denoting the data set in the book by $\chi_0$, they showed that the least squared estimator, $\hat{\boldsymbol{\theta}}(\chi_o)$, of $\boldsymbol{\theta}(F)$ is $(34.17, -0.057)^T$.

Assume further that the hormone data set in the book was created by linking the $y$ variables in data set A, with the $x$ variables in data set B, each with 27 records, using four linking variables e.g. name, sex, age and address. Due to linking errors in the four linking variables, the researcher obtains data set $\chi$, the incorrectly constituted data set, and $\hat{\boldsymbol{\theta}}(\chi),$ which are subject to linking errors. In the real world, $\chi_o$ is not known. Our objective here is to "recover" an unbiased estimator of $\boldsymbol{\theta}(F)$ from $\chi,$ by using the iterative bootstrap to correct $\hat{\boldsymbol{\theta}}(\chi)$ for linkage bias.

For linkage bias correction, the researcher needs some basic information about the linkage process. It can either be the linkage probabilities, $\hat{m}_l$, and $\hat{u}_l, l = 1,..,4,$ or the agreement matrix $\Gamma$ of dimension 729 (=27*27) * 4 (from which the linkage probabilities can be estimated using the EM algorithm). For the purpose of this example, we assume that the linkage probabilities are available to the researcher.

For this experiment we constructed an incorrectly linked data set of hormone data, $\chi$, as follows

1) Create the agreement matrix, $\Gamma$, using $(\hat{m}_1, \hat{m}_2, \hat{m}_3, \hat{m}_4) = (0.81, 0.62, 0.75, 0.83)$ and $(\hat{u}_1, \hat{u}_2, \hat{u}_3, \hat{u}_4) = (0.17, 0.19, 0.15, 0.25)$;
2) Calculate the FS weight for each of the 729 record pairs; and
3) Choose the 27 record pairs with the highest FS weights, subject to the 1-1 linking constraint.

These 27 record pairs, given in Appendix 1, is the incorrectly constituted $\chi$. Standard calculations show that $\hat{\theta}(\chi) = (31.55, -0.042)^T$. The bootstrapping results using AMBIS are summarized in Table 1. In this example, we chose $\eta_0 = 0.5$. We provide the true parameter values, biased estimates, and bias corrected estimates with corresponding 95% Percentile interval for the standard significance test, and residual bias estimates. The results indicated that there is no detectable reduction in bias beyond the triple and double bootstrap for the intercept and slope coefficients respectively.

Table 1: Removing linking bias results for the regression coefficients of the hormone data

| Coefficient | Intercept | Slope (Hours) |
|---|---|---|
| **Biased Free Estimate** | 34.17 | -0.057 |
| **Biased Estimate** | 31.55 | -0.042 |
| $k_i$ | 4 | 3 |
| $B_{i,k,\max}$ | 145 | 331 |
| $\Delta(\tilde{\theta}_i^{(k)})$ | -0.50 | -0.0014 |
| **95 Percentile Interval of** $\Delta(\tilde{\theta}_i^{(k)})$ | (-2.23, 0.22) | (-0.0058, 0.0002) |
| $\tilde{\theta}_i^{(k_i-1)}$ | 33.16 | -0.058 |
| **95 Percentile Interval of** $\tilde{\theta}_i^{(k_i-1)}$ | (30.84, 35.42) | (-0.064, -0.053) |

## 4.2 Empirical Application – Australian transition to employment, 2008 - 2012

We aim at estimating the odds of employment in 2012. The outcome variable is the probability of transitioning from being not employed (i.e. either unemployed or not in the labor force) in 2008 to being employed in 2012. The model includes a number of covariates measured in 2012 as well as the labor force status of the individual in 2008. The selected auxiliary contemporaneous variables include whether the individual resides in the capital city vs. balance of state in 2012, their age, sex, birthplace, labour force status, highest educational attainment, highest year of school completed, highest non-school qualifications and survey weight.

The data are from the 2008 and 2012 Labour Mobility surveys (ABS, 2008)(ABS, 2012) for the state of Tasmania in Australia, which are accessible for download by authorized users through the Australian Bureau of Statistics (ABS) website. Each survey provides unit record data for individuals aged 15 and over who had worked at some point during the year ending in February 2008 (33,231 records) or 2012 (32,119 records), which constitutes over 1,067 million record pairs. Based on the surveys' sampling fraction, we estimated that there are only 1,446 matched pairs. As there is no published information on the specific individuals included in the surveys, we employed probability matching to link records from the 2012 data to corresponding records from 2008. Using the EM algorithm, the $m$ and $u$ probabilities for the linking variables are estimated as: $\boldsymbol{m}$ = (0.62, 0.65, 0.9, 0.9, 0.56, 0.56, 0.66)$^T$ and $\boldsymbol{u}$ = (0.30, 0.28, 0.11, 0.12, 0.09, 0.27, 0.38)$^T$ rounded to second decimal place. To illustrate the methods of this paper, we took a 10% sample of the 1,446 matched pairs as our biased integrated data set.

Given the large number of covariates available in the file, we used LASSO to eliminate those not useful for the analysis. With an optimal regularization parameter of 0.0038, the selected covariates for 2012 are capital city/balance of state, age, sex, and survey weight. These variables were chosen as auxiliary predictors for the logistic regression model used to estimate the log odds of employment in 2012. In addition, the 2008 labor force status is a covariate in the final model, given that we aim at estimating the transition probability. In this example, we adopt a superpopulation view under an informative sampling design; both target variables and auxiliary are subject to variation due to sampling. Because respondents in the 2012 survey were selected with unequal selection probabilities, and thus the sampling design is informative (Pfeffermann et al., 1998). For the case of informative sampling, the sample distribution differs from the population distribution, and applying a standard logistic regression model without adjustment can result in biased regression coefficients. To address this issue, Pfeffermann (1996) proposed several methods, including incorporating the survey weight as a covariate in the model. We include the logaritm of the survey weights (Skinner, 1994) as a control variable in the logistic regression model to adress the effects of informative samples for regression analysis.

The logistic regression equation used in this exercise is:

$$log\left(\frac{p}{1-p}\right) = \alpha + \beta_1 \cdot LFS08 + \beta_2 \cdot CapCity\_BalanceOfState + \beta_3 \cdot Age + \beta_4 \cdot Sex + \beta_5 \cdot Log(SurveyWeight)$$

where $LFS08$ is a binary variable for the person's labour force status in 2008. The are equal to one if the person was employed and 0 otherwise; and $p$ represents the probability of being in employment in 2012. Table 2 gives the estimated coefficients based on the 10% pairs linked using the Fellegi-Sunter algorithm for Tasmania (Tas) before and after linkage bias removal.

Table 2: Estimated logistic regression coefficients for the transition to employment from 2008 to 2012 model for Tasmania (Australia)

| Coefficient | Slope 1 | Slope 2 | Slope 3 | Slope 4 | Slope 5 |
|---|---|---|---|---|---|
| **Biased Estimate** | 1.96 | -0.08 | 0.01 | -0.47 | 0.27 |
| $k_i$ | 3 | 3 | 3 | 2 | 4 |
| $B_{i,k,\max}$ | 2,567 | 2,218 | 2,726 | 169 | 201 |
| $\Delta(\tilde{\theta}_i^{(k)})$ | 0.0050 | -0.0001 | 0 | -0.0029 | -0.0061 |
| 95 Percentile Interval of $\Delta(\tilde{\theta}_i^{(k)})$ | (-0.033, 0.005) | (-0.02, 0.01) | (-0.001, 0.001) | (-0.023, 0.034) | (-0.29, 0.16) |
| $\tilde{\theta}_i^{(k_i-1)}$ | 2.28 | -0.04 | 0.02 | -0.50 | -0.015 |
| 95 Percentile | (2.21, 2.36) | (-0.06, 0.021) | (0.017, 0.021) | (-0.55, -0.45) | (-0.33, 0.24) |

| Interval of $\tilde{\theta}_i^{(k_i-1)}$ | | -0.02) | | | |
|---|---|---|---|---|---|
| Odds Ratio (with linkage errors) | 7.1 | 0.92 | 1.01 | 0.63 | 1.31 |
| Odds Ratio (without linkage errors) | 9.8 | 0.96 | 1.02 | 0.65 | 0.99 |

**Legend**: Slope 1 = *LFS08*; Slope 2 = *CapCity_BalanceOfState*; Slope 3 = *Age*; Slope 4 = *Sex*; Slope 5 = *log(SurveyWeight1)*

Table 2 shows that, holding other covariates constant, prior employment is the dominant predictor of being employed in 2012. Individuals employed in 2008 have about 10 times the odds of 2012 employment (OR = 9.8; 95% CI on OR ≈ 9.1–10.6). Sex also matters: men have ~39% lower odds than women (OR = 0.61; 95% CI ≈ 0.58–0.64). Location has a small effect: living in the capital city vs balance-of-state is associated with ~4% lower odds (OR = 0.96; 95% CI ≈ 0.95–0.97). Age has a modest positive effect, about +2% per year (OR = 1.02; 95% CI ≈ 1.018–1.021).

Estimates before correcting for linkage error attenuate the LFS08 effect to about 7 times the odds. After eliminating statistically significant linkage errors through record linkage refinement, the estimated effect rises to about 10 times, underscoring the value of integrating the 2008 and 2012 datasets with linkage-error correction.

## 5. Conclusion

In this paper, we provide a method to construct linkage bias-corrected estimators for the parameters of models estimated from linked data sets. Due to error in the decision that a pair is a match, a bias arises when estimating the parameters of an unknown distribution, $F$. The key in applying AMBoP sampling is to consider the agreement matrix, $\Gamma$, as a sample from two multivariate Bernoulli distributions, $\Pi_{i=1}^{L} Bern(\hat{m}_i)$ and $\Pi_{i=1}^{L} Bern(\hat{u}_i)$, and create bootstraps of $\Gamma$ by sampling independently and repeatedly from both distributions. Each bootstrap, together with the FS algorithm, will enable the creation of a bootstrap linked data set, i.e. a AMBIS, from which a bootstrap estimate of the parameter of interest can be computed. In addition, we also use OR sampling to improve the efficiency for creating bootstrap samples to compute the percentile confidence intervals for the bias corrected regression coefficients.

In this paper, we have demonstrated the efficacy of the methods by applying them to bias correction for estimating regression coefficients of a simulated linked data set, and the logistic regression coefficients for the linked data set between the 2008 and 2012 labour mobility surveys for the state of Tasmania from the Australian Bureau of Statistics.

Finally, so that the correction methods in this paper can be applied, the researcher needs to have access to either the agreement matrix, $\Gamma$, or the estimates of the probability of matched and unmatched pairs, $\hat{m}$ and $\hat{u}$, used by the data integrator, e.g. a national statistics office, to construct the integrated data set. Currently this information is not published by the data integrator. We hope that the methods outlined in this paper, which can provide linkage bias-corrected estimators from integrated data, are sufficiently convincing for data integrators to change their current publication practice, and induce the publication of the required information to enable researchers to carry out linkage bias correction.

**Acknowledgement**

We would like to thank the referees for their comments which led to substantial methodological improvements of the paper.

## Appendix 1 – The Hormone Data Sets

The correctly constituted data set (27 pairs) – sourced from Efron and Tibshirani (1993).

| hrs | amount | hrs | amount | hrs | amount |
|-----|--------|-----|--------|-----|--------|
| 99  | 25.8   | 376 | 16.3   | 119 | 28.8   |
| 152 | 20.5   | 385 | 11.6   | 188 | 22.0   |
| 293 | 14.3   | 402 | 11.8   | 115 | 29.7   |
| 155 | 23.2   | 29  | 32.5   | 88  | 28.9   |
| 196 | 20.6   | 76  | 32.0   | 58  | 32.8   |
| 53  | 31.1   | 296 | 18.0   | 49  | 32.5   |
| 184 | 20.9   | 151 | 24.1   | 150 | 25.4   |
| 171 | 20.9   | 177 | 26.5   | 107 | 31.7   |
| 52  | 30.4   | 209 | 25.8   | 125 | 28.5   |

$$\hat{\theta}(\chi_0) = (\hat{\beta}_0, \hat{\beta}_1)^T = (34.17, -0.057)^T$$

The incorrectly constituted data set (10 false positive pairs) re-created from the above set using:

$$(\hat{m}_1, \hat{m}_2, \hat{m}_3, \hat{m}_4) = (0.81, 0.62, 0.75, 0.83); \quad (\hat{u}_1, \hat{u}_2, \hat{u}_3, \hat{u}_4) = (0.17, 0.19, 0.15, 0.25)$$

| hrs | amount | hrs | amount | hrs | amount |
|-----|--------|-----|--------|-----|--------|
| 99  | 32.5   | 376 | 16.3   | 119 | 28.8   |
| 152 | 20.5   | 385 | 11.6   | 188 | 22.0   |
| 293 | 14.3   | 402 | 25.4   | 115 | 29.7   |
| 155 | 23.2   | 29  | 31.7   | 88  | 20.9   |
| 196 | 20.6   | 76  | 32.8   | 58  | 25.8   |
| 53  | 31.1   | 296 | 18.0   | 49  | 32.5   |
| 184 | 20.9   | 151 | 24.1   | 150 | 32.0   |
| 171 | 11.8   | 177 | 26.5   | 107 | 25.8   |
| 52  | 30.4   | 209 | 28.9   | 125 | 28.5   |

$$\hat{\theta}(\chi) = (\hat{\beta}_0, \hat{\beta}_1)^T = (31.55, -0.042)^T$$